**Enhancing Malware Detection with Generative AI: Using Variational Autoencoders to Boost Machine Learning Classifiers' Performance**

Mohammad Alharbi
Department of Computer Science
North Dakota State University
1320 Albrecht Blvd., Room 258
Fargo, ND 58108
Phone: +1-701-231-8562
Fax: +1-701-231-8255
Email: mohammad.k.alharbi@ndsu.edu

Jeremy Straub
Center for Cybersecurity & AI
University of West Florida
220 W. Garden St., Suite 250
Pensacola, FL 32502
Phone: +1-850-474-2999
Email: jstraub@uwf.edu

**Abstract**

The advancement of malware poses obstacles for cybersecurity, necessitating the development of advanced detection techniques. This paper proposes an approach to enhance malware detection through the use of a generative artificial intelligence model. Specifically, variational autoencoders (VAEs) are used with the random forest, XGBoost and sequential model machine learning classifiers. Generated synthetic malware samples are used to address the critical issue of data scarcity for new or less common malware types. This approach can be used to augment datasets to improve classifier robustness.

The proposed methodology uses VAEs to create high-quality diverse synthetic datasets that closely mimic real-world malware data. The effectiveness of these augmented datasets is evaluated by comparing the performance of the machine learning classifiers when they are trained with the original data and when they are trained with the synthetic data-augmented datasets. The results demonstrate a notable improvement in the accuracy, precision, recall and F1-scores of the classifiers, when they are trained using the augmented datasets. The enhanced performance for detecting various malware classes shows the potential of this approach to facilitate adaptation to evolving malware threats effectively. This work demonstrates the utility of generative AI for cybersecurity. It also provides a foundation for future research aimed at developing more resilient and adaptive malware detection systems.

**Keywords:** malware detection, generative artificial intelligence, XGBoost, RFC, sequential, variational autoencoders (VAEs)

## 1. Introduction

Malware presents a threat to both individuals and organizations. It has a history tracing back to the 1970s and the emergence of the creeper virus [1]. The subsequent decades have seen a surge of malware, encompassing hundreds of thousands of variants, each designed to steal data and inflict disruption and damage [1].

Despite advancements in cybersecurity, malware continues to pose a formidable challenge. A 2021 study indicated that over 60% of organizations are still vulnerable to malware threats and – alarmingly – only approximately 5% of company documents are adequately secured [1]. The financial and temporal costs associated with these threats are substantial. Patil, et al. reported that the average time required to recover from a malware attack is approximately 50 days [1]. The average cost of cybersecurity incidents, including malware related attacks, reached approximately $4.44 million in 2025, according to an IBM study [2].

In light of these factors, there is a need for new antimalware methodologies. Machine learning – and particularly generative AI – may have a key role to play in enhancing malware detection capabilities. Generative artificial intelligence has been shown to be adept at creating synthetic data that closely mirrors real world data [3]. This offers a promising avenue for training models with synthetic samples. It may help with the challenge, in the domain of malware detection, of the scarcity of labeled samples for new or less common malware families.

This approach is particularly beneficial, for malware detection applications, due to the challenges presented by trying to acquire large, sufficiently diverse, and high-quality real-world malware datasets. The full implications of training models primarily or exclusively with synthetic data remain an area of ongoing research [3]. A notable concern is the phenomenon of model autophagy disorder (MAD), where generative models may degrade in quality over time, without a regular influx of new real data [4]. This highlights the need to use synthetic data carefully as a supplement to real data, rather than as a full replacement [4]. Although generative models have shown success in several domains, their application to structured cybersecurity data requires careful design [3].

VAEs are used for malware data augmentation because they can learn complex data distributions and generate new samples that are similar to the original data [5]. They operate by learning a latent representation of the input data, which consists of known malware samples. They achieve this using an encoder/decoder structure. The encoder compresses the input data into a latent, lower-dimensional space and the decoder reconstructs the data back to its original higher-dimensional space form, from this latent space. The key capability of VAEs, for this process, is their ability to model the probability distribution of the input data in the latent space. By sampling from this distribution, VAEs can generate new data samples that are similar to – but not exact replicas of – the original data. This capability allows VAEs to generate diverse malware samples which can be used to augment existing datasets and enhance the robustness of classifiers.

Generative AI models, especially VAEs, can be used for generating adversarial samples. VAEs can produce samples that can be used to train and improve the detection performance of malware detection systems. These samples can mimic potential mutations or evolutions of malware, providing a proactive approach to developing more effective defense mechanisms. This use of VAEs contributes to the robustness of malware classifiers and aids them with staying up to date with evolving cyber threats.

Tabular data with heterogeneous relationships, limited sample sizes and the need for domain-specific knowledge poses distinct hurdles [6]. Yet, the potential value of generating synthetic samples for this data type is immense, particularly in application areas where data is scarce [6]. This paper, thus, explores the application of using generative AI, specifically VAEs, to generate synthetic malware samples.

VAEs are used, herein, to create synthetic datasets, based on malware data that has been previously used to evaluate classifier performance in [7]. In [7], Alenezi and Ludwig improved the explainability of malware classification, as compared to [8] and [9], which didn't substantively consider this topic; however, the methods that were used to do this underperformed the approaches presented in [8] and [9] in terms of classification accuracy. The use of VAEs to create synthetic data, herein, results in improved performance

– in terms of the key accuracy, precision, recall and F1-score metrics – approaching the results reported in [8] and [9], while still providing the same explainability benefits, by using the same classifier techniques, discussed in [7].

The synthetic data is used to augment the existing training data. Then, the same classification models, that were used previously in [7], are used to evaluate the efficacy of the newly generated synthetic data's use in enhancing the models' training. This allows for a direct comparison between the performance of the classification models trained with the original real-world data and their efficacy when trained with a data set augmented by the synthetic data generated by the VAEs. This approach assesses the benefits of using synthetic data in malware classification and provides insights into the effectiveness and limitations of using generative AI models for replicating complex data structures for cybersecurity applications.

This paper continues with Section 2, which presents prior related work relevant to the utilization of generative AI models for the generation of synthetic samples that are aimed at enhancing threat detection through improved classification. In Section 3, the proposed technique is detailed. Background regarding using VAEs for generating synthetic malware data and a brief background of the ML models – the random forest classifier, XGboost classifier and the sequential model – are also presented. Section 4 presents the experimentation that was conducted and its results. The dataset that was used is also described and the outputs of the generative AI model that was used are presented and compared to the results from [7]. Next, in Section 5, the proposed data augmentation approach is compared to the approaches proposed by Mahdavifar, et al., in [8, 9], and another recent study also using this dataset. Mahdavifar, et al. compiled the dataset that is used for the analysis in this paper. Then, Section 6 compares the performance of the approach presented in this paper with two other synthetic data-based approaches. Finally, Section 7 presents the key findings and conclusions drawn from this research.

## 2. Related Work

ML algorithms have been previously developed for malware detection and classification. Prior studies [10, 11] have shown that employing ML for malware detection can be beneficial; however, despite technological advancements, challenges remain.

Generative AI has been used in the field of data synthesis for complex domains such as complex tabular data [6]. This introduces a set of challenges which are different from those in synthesizing visual or textual data. One challenge lies in the heterogeneity inherent in tabular data, which can include a mix of categorical, numerical, and other data types within a single dataset. Addressing this diversity requires modeling techniques capable of managing these data types and their interrelationships. It is crucial for models to learn and utilize the underlying relational structures to capture the complex dependencies between variables. The integration of domain-specific knowledge is also a challenge. Compounding these challenges are issues such as limited sample sizes and data sparsity, which can impede the training of generative models. Accurately labeled training datasets are needed to allow models to attain high levels of accuracy in malware identification. All of this is further complicated by the frequent emergence of new malware variants [12].

Saad, et al. [10] reviewed the challenges of applying ML to malware detection, focusing on issues such as evolving malware behaviors, high retraining costs, and model interpretability. The study highlighted the potential of combining dynamic and behavioral analysis with machine learning to address advanced threats like fileless and polymorphic malware. While classifiers – such as random forest – have shown effectiveness, the need for frequent updates – to adapt to the rapidly changing threat landscape – was discussed. They also stressed the importance of using interpretable models to build trust among cybersecurity analysts, noting this is a persistent challenge in the field.

Patil, et al. [1] proposed a framework to improve the robustness of ML-based malware detection using adversarial training. The study highlighted the susceptibility of AI models to adversarial attacks, where attackers exploit model weaknesses by introducing carefully crafted perturbations. By generating adversarial malware samples and retraining models, such as random forest and support vector machines, they achieved notable improvements in detection accuracy and resilience. However, the study also identified challenges, including computational complexity and the increasing sophistication of adversarial attacks, showing the need for the ongoing adaptation of malware detection systems.

Recent research has also explored generative models that can produce complex data by identifying the dependencies between variables. VAEs, built on neural networks and optimized through stochastic gradient descent, have emerged as a promising solution for this. By learning the underlying data distributions, VAEs generate samples that closely resemble the data in the original dataset. This capability is important in domains such as cybersecurity, where preserving the relationships among features is essential because the attributes in tabular data are often correlated. Maintaining these correlations is critical to preserve the utility of the data for downstream classification tasks [6].

Wan, et al. [5] introduced a VAE-based approach to generating synthetic data for addressing imbalanced learning challenges. Unlike traditional methods, such as the synthetic minority oversampling technique (SMOTE), which rely on Euclidean distance and often struggle with high-dimensional data, VAEs learn the underlying data distribution to produce realistic and diverse synthetic samples. This method demonstrated significant improvements in classification performance with imbalanced datasets, showing its potential for applications in areas like cybersecurity, where data scarcity and imbalance are prevalent issues. However, the quality of VAE-generated data was shown to be highly dependent on model architecture and appropriate hyperparameter tuning, highlighting the need for careful implementation.

Fajardo, et al. [13] investigated the use of VAEs to address class imbalance in datasets by generating synthetic samples for minority classes. The study demonstrated that using VAEs, which learn class-specific data distributions, significantly improved the performance of downstream classifiers, such as random forest and XGBoost. Compared to traditional oversampling techniques, like SMOTE, VAEs were shown to achieve more robust performance gains, particularly in scenarios with severe class imbalance. However, challenges – including computational complexity and the need to ensure the quality of synthetic data – were identified, again highlighting the importance of careful parameter tuning and architecture selection for effective VAE implementation for real-world applications.

Faisal, et al. [12] conducted a survey of artificial intelligence techniques for malware detection, focusing on both ML and deep learning approaches. The study highlighted the effectiveness of models, such as random forest and XGBoost, for classifying malware in structured datasets. The suitability of deep learning methods, including sequential and recurrent neural networks, for analyzing unstructured and evolving malware behaviors was also demonstrated. They also identified several key challenges, including dataset availability, model scalability, and the growing complexity of advanced threats, like polymorphic malware. These findings suggest the potential effectiveness of integrating classifiers, like random forest, with deep learning models to develop robust malware detection systems.

Bortoli, Duarte and Guerreiro [14] used the dynamic execution characteristics from the CICMalDroid2020 dataset and compared 24 machine-learning configurations using the naive Bayes, SVM/SVC, random forest, and XGBoost models. The preprocessing techniques that were used included variance filtering, scaling, PCA, truncated SVD, and F-test or Chi-square feature selection. Each configuration was trained with and without SMOTE, using Optuna hyperparameter optimization and nested cross-validation. Bortoli, Duarte and Guerreiro used weighted recall as their evaluative metric.

Othman, et al. [15] used the dynamic features extracted from the CCCS-CIC-AndMal2020 dataset, applied mutual-information feature weighting and MinMax normalization, and trained a conditional GAN (CGAN) to generate synthetic malware samples. They evaluated two CGAN conditioning strategies: multi-class generation, using all 12 malware classes at once, and binary-class generation, using smaller class pairs. Synthetic data was merged with the original data and evaluated using the LightGBM and random forest techniques. They also tested histogram-matching post-processing.

## 3. Methodology

The section presents the details regarding how VAEs were used to generate synthetic malware data in this study. The process – beginning with the preparation of the original malware dataset, followed by the generation of synthetic data using VAEs, and concluding with the use of various classification models – is described in this section. Information regarding the implementation of the machine learning models and the random forest, XGboost and sequential model classifiers is also included.

### *3.1. Study Process*

This section presents the process used for this study. It is depicted in Figure 1 and each step is now discussed.

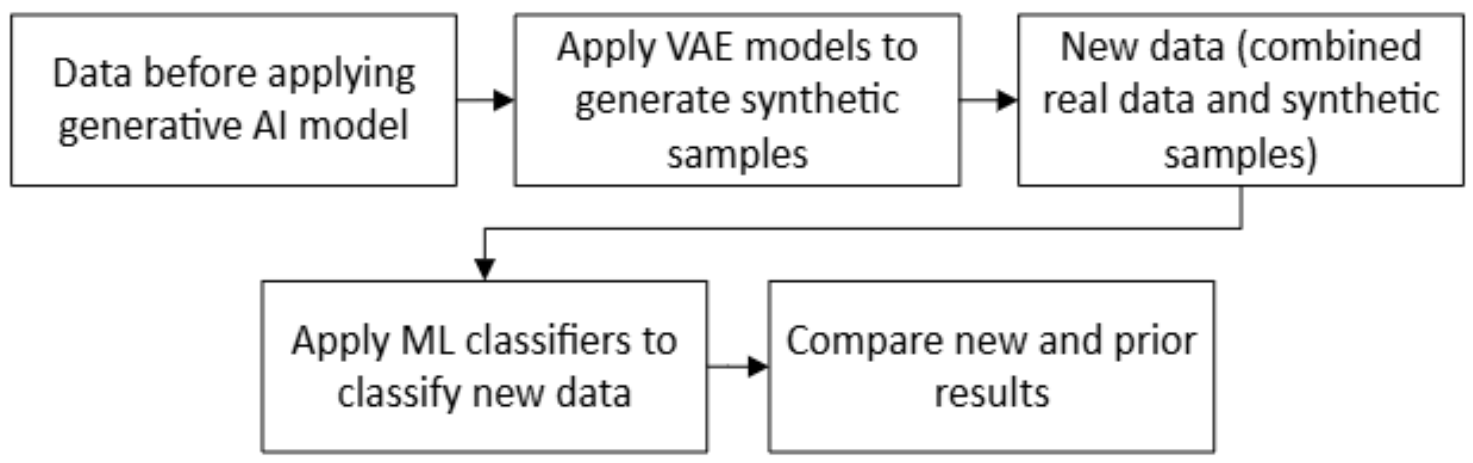

**Figure 1.** Study Process Workflow.

The process starts with the original data, before applying the generative AI model. Then, VAEs are used to generate synthetic samples. This process creates the data that will be used to augment the original dataset with artificially created data points which are likely to improve the model's training and, thus, performance.

Third, the newly generated synthetic samples from the VAE model are combined with the original dataset. This merged dataset is designed to provide a more robust and varied set of data for machine learning.

Fourth, the combined dataset is used to train the machine learning models. Specifically, the deep learning, XGboost, and random forest techniques are used.

Finally, the performance of the proposed approach is evaluated. The results are compared with those found in prior work (e.g., [7]).

This is a typical machine learning pipeline. The data is augmented, then used to train models, data is presented for classification for testing purposes, and the outcomes evaluated against a benchmark.

### *3.2. Variational Autoencoders*

VAEs are a key innovation for generative modeling. They have attracted substantial interest because of their ability to create synthetic data [16]. Unlike traditional autoencoders, VAEs employ a probabilistic method for data generation and learn to map data into a latent space. This allows for the production of new data points that conform to the original data's probabilistic distribution [16]. This feature is beneficial in areas like cybersecurity and malware analysis, where data can be scarce and augmenting data is challenging. VAEs' capacity to replicate complex and high-dimensional data distributions makes them useful for supplementing datasets, especially for cases where gathering additional real-world data is either not feasible or otherwise problematic [17]. The use of VAEs for generating synthetic data for malware detection is presented and evaluated herein. It offers a way to improve the strength and accuracy of malware classification models.

### *3.3. Random Forest Classifier*

The random forest classifier is a method where random samples are drawn from a dataset to construct decision trees. Each tree generates a prediction result, and the final prediction is determined by the most common result among the trees [18].

The random forest classifier is known for its effectiveness for classification tasks. It uses a set of parameters that influence its performance and decision-making process. One key parameter is the criterion that is used for measuring the quality of a split in the decision trees in the random forest [19]. Two commonly used criteria are the Gini index and entropy.

The Gini index evaluates the purity of a node in a decision tree. It is calculated using the equation [7]:

$$Gini = 1 - \sum_{i=1}^{C} P_i^2 \tag{1}$$

where $P_i$ represents the relative frequency of class i observed in the dataset and C is the total number of classes. The Gini index determines how nodes in the decision tree are divided. The probability and class are assessed to compute the Gini value for each branch. This value indicates which branch of a node is more likely to have uniform (i.e., all branch members belonging to a single class) class membership, based on the class probabilities [7].

An alternative approach is the use of entropy, which measures the level of uncertainty or impurity in a dataset. Higher entropy indicates more mixed class distributions, while lower entropy indicates more homogeneous data. In decision tree construction, entropy is used to determine how nodes are split based on class probabilities, where the probability of each class is calculated at a node and used to evaluate how the data is distributed across classes. [7].

Entropy is calculated using the following equation [7]:

$$Entropy = \sum_{i=1}^{C} -P_i \times log_2 P_i \tag{2}$$

Entropy also relies on probabilities, but it has a different range, as compared to the Gini index. While the Gini index's values range from 0 to 0.5, indicating varying levels of node purity, entropy values span from 0 to 1 [20]. The range difference can be a factor when selecting the most suitable criterion for a given task. The Gini index is often preferred due to its narrower range [20]. The selection of a set of criteria, such as the Gini index or entropy, is crucial for steering the random forest classifier towards making accurate predictions through the effective selection of features and the efficient division of nodes within the decision trees. The criteria guide the random forest classifier, ensuring that it accurately classifies data by enhancing the organization and analysis of decision trees. Other key parameters include 'max depth', which determines the tree's depth, and 'n estimators', which specifies the forest's tree count.

For the experimentation presented herein, the Gini criterion was used along with a ‘max depth’ value of 30 and an ‘n estimaotrs’ value of 200.

### *3.4. XGboost*

XGboost classification is an ensemble ML technique that is a form of gradient boosting. It provides rapid processing and high scalability [7, 21]. Extreme gradient boosting incorporates both gradient boosting elements and specialized features, such as multiclass classification [25] and histogram-based tree construction [26]. It uses second-order gradients and incorporates regularization techniques to achieve more precise approximations [7, 21].

XGboost first calculates initial prediction values and error margins. Following this, a model is developed using independent variables and residual errors [7, 22]. This helps to fine-tune its predictions [22]. This process involves three key parameters, that were identified in [7]:

1) “binary: logistic” – This setting configures XGboost to perform “logistic regression for binary classification”. This makes the model output a probability score for each sample, which is then used to assign the final class label [7, 25].
2) “max_depth” – This setting controls the maximum depth of each decision tree, limiting the complexity of the model [25].
3) “n_estimators” – This setting configures the number of rounds (e.g., the number of trees added sequentially) of the boosting process [21, 27].

By adjusting these parameters, the model’s performance can be improved and complexity and overfitting can be controlled. This paper used the same XGboost settings as [7] (i.e., max_depth=10 and n_estimators=800).

### *3.5. The Sequential Model*

A sequential model, functioning as a classifier, consists of a linear arrangement of layers. Every layer in this structure has an input tensor and output tensor. The model’s first layer uses the rectified linear unit (ReLU) activation function, represented by the equation [7]:

$$F(x) = (0, x) \tag{3}$$

where $F(x)$ denotes the mapping function and $x$ is the input variable.

The final layer employs the softmax activation function, to extend the capabilities of the first layer [23]. This is shown in the equation [7]:

$$Softmax(x_i) = \frac{e^{x_i}}{\sum_i e^{x_i}} \tag{4}$$

where $x_i$ is the input (value) for class $i$, and $\sum_i e^{x_i}$ is the sum of the exponentiated input values across all classes. The softmax function converts these values into probabilities that sum to 1.

Following the establishment of the network architecture, key elements – such as the loss function, optimizer, and metrics for prediction – are specified [7].

The settings used with the sequential model for the experimentation presented herein are listed in Table 1.

**Table 1.** Sequential Model Settings.

| Setting | Value |
|---|---|
| Hidden layers | 3 dense layers |
| Neurons per hidden layer | 256, 128, 64 |
| Hidden-layer activation | ReLU |
| Output activation | Softmax |
| Output layer | Dense layer with num_classes units |
| Dropout | 0.2 after each hidden layer |
| Optimizer | Adam |
| Learning rate | 0.0005 |
| Early stopping | Monitor val_loss, patience = 15, restore best weights |

## 4. Experimentation and Results

This section presents the experimentation that was conducted and its results. It starts by describing the dataset that was used. This is followed by a discussion of the outputs of the generative AI model that was used to generate the synthetic samples and the ML models that were trained and evaluated, after generating the new samples. These results are then compared to the results from prior studies.

### *4.1. Data Set*

The data set that was used was provided by the Canadian Institute for Cybersecurity at the University of New Brunswick [24]. It is called "CICMalDroid 2020" and contains 11,598 Android application samples that are represented by 469 extracted features and one class-label column [8]. In this study, only the extracted feature data were used. The class-label column was used as the label for training.

### *4.2. Synthetic Data Generation Using VAEs*

The VAE generative model [5], unlike traditional generative models that require meeting robust assumptions about data structure or depend on resource intensive inference methods, operates with minimal data requirements. Its training process utilizes backpropagation for rapid execution [5].

The probability of each data sample, $X$, is calculated with the equation [5]:

$$P(X) = \int P(X|z;\theta)P(z)dz \tag{5}$$

where $X$ denotes a data sample, $z$ is the latent variable, $P(z)$ is the prior distribution from which $z$ is sampled, $P(X \mid z;\theta)$ is the conditional distribution used to generate $X$ given $z$, and $P(X)$ is the probability of the sample $X$, computed by integrating over possible values of $z$ [5]. The parameter $\theta$ represents the model encoder and decoder weights.

In this study, a VAE was implemented using TensorFlow and Keras to generate synthetic data for data augmentation purposes. The process begins with the loading and preprocessing of the data, including the normalization of the feature values using StandardScaler and the separation of the target labels. The dataset is then divided into the training and testing sets.

The VAE model was configured with parameters for the input, latent, and intermediate layer dimensions. The encoder of the VAE is designed to map the input data to two latent space parameters: mean and log variance, which are used to generate latent representations. A sampling function was then employed to sample from this latent space. The decoder component then mapped these sampled latent points back to the original data space, producing reconstructed synthetic data points. The loss function of the VAE combines the reconstruction loss and Kullback-Leibler divergence, adjusted by respective weights, to balance the components. For synthetic data generation, the learned latent representations were perturbed and passed through the decoder to generate new samples, which were formatted and integrated with the original dataset. The augmented dataset was then saved as a new CSV file for subsequent use.

This approach increases the size of the training dataset. It also introduces variability, potentially improving the robustness and generalizability of the models trained with it. The VAE maintains proportionality to the original dataset's distribution.

The synthetic samples simulate new types of similar malware, providing the models with a broader spectrum of data for training. By integrating the synthetic data with the original dataset, the dataset was expanded, from the initial 11,598 records to have 14,096 records. This increased the volume of data available for training and diversified the data. This diversity allows the training to produce more robust and effective machine learning models and enhances their capability to detect a wider array of malware, including new variants.

Each model was trained using the resulting augmented training dataset.

### *4.3. Precision, Recall and F1-Score Results*

The models were evaluated using the test set from the original "CICMalDroid 2020" data. Three ML classifiers – random forest, Xgboost and the sequential model – were used, and the results were compared to the results presented in [7]. The precision, recall and F1-score results are presented in this section. Correct classification and accuracy results are presented subsequently.

Tables 2 through 4 present the evaluation of the malware detection models using the metrics of precision, recall, and F1 score. These metrices are presented by classification for the adware, banking malware, SMS malware, riskware, and benign software classes.

**Table 2.** Precision, Recall, and F1-Score Results from the Sequential Model Classifier.

| Classes | Using AVEs (This Study) | | | Results in [7] | | |
|---|---|---|---|---|---|---|
| | Precision | Recall | F1-Score | Precision | Recall | F1-Score |
| Adware | 0.7806 | 0.8282 | 0.8037 | 0.70 | 0.80 | 0.75 |
| Banking malware | 0.8303 | 0.8538 | 0.8419 | 0.82 | 0.84 | 0.83 |
| SMS malware | 0.9332 | 0.9771 | 0.9546 | 0.92 | 0.97 | 0.94 |
| Riskware | 0.9176 | 0.8460 | 0.8803 | 0.85 | 0.80 | 0.82 |
| Benign | 0.8851 | 0.8190 | 0.8507 | 0.88 | 0.73 | 0.80 |

**Table 3.** Precision, Recall, and F1-Score Results from the Random Forrest Classifier.

| Classes | Using AVEs (This Study) | | | Results in [7] | | |
|---|---|---|---|---|---|---|
| | Precision | Recall | F1-Score | Precision | Recall | F1-Score |
| Adware | 0.9066 | 0.9433 | 0.9246 | 0.85 | 0.93 | 0.89 |
| Banking malware | 0.9684 | 0.9130 | 0.9399 | 0.96 | 0.89 | 0.92 |
| SMS malware | 0.9729 | 0.9934 | 0.9831 | 0.98 | 0.99 | 0.98 |
| Riskware | 0.9579 | 0.9391 | 0.9484 | 0.94 | 0.91 | 0.92 |
| Benign | 0.9202 | 0.9428 | 0.9314 | 0.90 | 0.94 | 0.92 |

**Table 4.** Precision, Recall and F1-Score Results from the XGBoost Classifier.

| Classes | Using AVEs (This Study) | | | Results in [7] | | |
|---|---|---|---|---|---|---|
| | Precision | Recall | F1-Score | Precision | Recall | F1-Score |
| Adware | 0.9130 | 0.9277 | 0.9203 | 0.86 | 0.92 | 0.89 |
| Banking malware | 0.9507 | 0.9257 | 0.9380 | 0.94 | 0.91 | 0.92 |
| SMS malware | 0.9823 | 0.9898 | 0.9860 | 0.99 | 0.99 | 0.99 |
| Riskware | 0.9559 | 0.9468 | 0.9513 | 0.96 | 0.93 | 0.94 |
| Benign | 0.9343 | 0.9506 | 0.9424 | 0.92 | 0.93 | 0.93 |

As compared to [7], the models exhibited superior performance in terms of all three metrics, across most classes. For instance, for the sequential model, there was a notable increase in precision for the adware (0.78 vs. 0.70) and riskware (0.92 vs. 0.85) classes. For the random forest model, the proposed approach outperformed in precision for the classes banking malware (0.97 vs. 0.96) and benign (0.92 vs. 0.90). Similarly, the XGboost model, using the proposed approach, achieved higher scores for precision and recall for adware (0.91 vs. 0.86 for precision and 0.93 vs. 0.92 for recall) and banking malware (0.95 vs. 0.94 for precision and 0.93 vs. 0.91 for recall).

These results demonstrate the benefits of the proposed approach. They also characterize the effectiveness of the approach in accurately classifying various malware types, as compared to [7]. Notably, the proposed VAE-based approach improved performance in many of the areas that the prior approach performed the worst in.

### *4.4. Comparison of ML Classifiers*

Table 5 presents a comparison of the correct classification count results of this study and the results presented in [7]. The higher correct classification rates achieved by the machine learning models – sequential model, random forest, and XGBoost – demonstrate the efficacy of the proposed approach.

**Table 5.** Correct Classification Counts for Each of the Three ML Models.

| ML Model | This Study (Using VAEs) | Results in [7] | Improvement |
|---|---|---|---|
| Sequential Model | 2055 | 1986 | 3.5% |
| Random Forest | 2211 | 2183 | 1.3% |
| XGBoost | 2218 | 2197 | 0.1% |

Notably, using VAEs produced improvement in classification accuracy using all three models. The most notable improvement was observed using the sequential model, where correct classifications were achieved for 2,055 samples out of 2,320 records, as compared to the 1,986 records that were correctly classified in [7]. The random forest model also showed an increase, correctly classifying 2,211 records with the proposed approach, as compared to 2,183 in [7]. Similarly, the XGBoost model also showed an increase. The proposed approach achieved correct classifications for 2,218 records compared to the 2,197 records that were correctly classified in [7]. These results demonstrate the efficacy of the synthetic data training technique that was employed in this study.

The VAE-based approach also demonstrated improvement in accuracy, as compared to [7]. This is shown in Table 6.

**Table 6.** Overall Accuracy of the Three ML Models.

| ML Model | This Study (Using VAEs) | Results in [7] |
|---|---|---|
| Sequential Model | 88.58% | 86% |
| Random Forest | 95.30% | 94% |
| XGBoost | 95.60% | 95% |

For the sequential model, the proposed VAE-based approach achieved an 89% accuracy rate, which is a 3% increase over [7]'s 86% accuracy. For the random forest model, the proposed approach's accuracy rate was 95%, outperforming [7] by 1%. For the XGBoost model, the proposed approach had a 96% accuracy rate, surpassing [7]'s 95% accuracy. These results demonstrate the effectiveness of the proposed technique.

A comparison of the accuracy of the proposed technique is also presented by malware class and classifier. This data is presented in Table 7.

**Table 7.** Accuracy of the Models by Malware Class.

| Classes | Sequential Model | Random Forest | XGBoost |
|---|---|---|---|
| Adware | 0.8282 | 0.9433 | 0.9277 |
| Banking malware | 0.8538 | 0.9130 | 0.9257 |
| SMS malware | 0.9771 | 0.9934 | 0.9898 |
| Riskware | 0.8460 | 0.9391 | 0.9468 |
| Benign | 0.8190 | 0.9428 | 0.9506 |

Random forest was shown to perform the best for the adware and SMS malware classes. XGBoost was shown to perform best for the banking malware, riskware and benign categories.

The limited improvement, as compared to other techniques, across all three models suggest that this approach has produced a slightly more refined and accurate ML model, thus enhancing performance.

## 5. Comparison to Other Approaches

This section compares the results generated from the approach presented herein to prior work by Alenezi and Ludwig [7] and Mahdavifar, et al. [8, 9]. Mahdavifar, et al. curated the "CICMalDroid 2020" dataset and used it to assess several techniques; however, they did not substantively consider the explainability of those techniques. Alenezi and Ludwig built on this, using models that could be interpreted using Shapley additive explanations (commonly known as SHAP); however, this resulted in performance – in terms of the accuracy precision, recall and F-1 score metrics – that was inferior to [8] and accuracy performance that was, by category, inferior to both [8] and [9].

The work presented herein has improved on the performance of the classifiers used in [7] by incorporating synthetic data, enhancing the performance of the classifiers to approach the levels reported by Mahdavifar, et al. in [8] and [9]. The random forest and XGBoost models, both herein and in [7], outperform [8] in terms of adware classification accuracy.

Table 8 presents this comparison, in terms of overall accuracy, precision, recall and F1-score. Comparison by malware class is presented in Table 9.

**Table 8.** Comparison to Other Approaches in Terms of Accuracy, Precision, Recall and F1-Score.

| Approach | Accuracy | Precision | Recall | F1-score |
|---|---|---|---|---|
| Sequential model (current) | 88.58 | 88.66 | 88.58 | 88.54 |
| Random forest (current) | 95.30 | 95.34 | 95.30 | 95.29 |
| XGBoost (current) | 95.60 | 95.61 | 95.60 | 95.60 |
| Sequential model (from [7]) | 86 | 86 | 86 | 86 |
| XGBoost (from [7]) | 95 | 95 | 95 | 95 |

| Random forest (from [7]) | 94 | 94 | 94 | 94 |
|---|---|---|---|---|
| Approach from [8] | 96.70 | 99.16 | 96.54 | 97.84 |

**Table 9.** Comparison to Other Approaches by Malware Class.

| **Classes** | **Sequential Model** | | **Random Forest** | | **XGBoost** | | **Mahdavifar, et al.** | |
|---|---|---|---|---|---|---|---|---|
| | Current | [7] | Current | [7] | Current | [7] | [8] | [9] |
| Adware | 0.8282 | 0.8049 | 0.9433 | 0.9309 | 0.9277 | 0.9228 | 0.85 | 0.96 |
| Banking malware | 0.8538 | 0.8397 | 0.9130 | 0.8923 | 0.9257 | 0.9115 | 0.96 | 0.96 |
| SMS malware | 0.9771 | 0.9670 | 0.9934 | 0.9878 | 0.9898 | 0.9914 | 1.00 | 0.99 |
| Riskware | 0.8460 | 0.7965 | 0.9391 | 0.9139 | 0.9468 | 0.9256 | 0.98 | 0.95 |
| Benign | 0.8190 | 0.7317 | 0.9428 | 0.9360 | 0.9506 | 0.9329 | 0.98 | 0.95 |

Abedin and Mehrub [28] also used the "CICMalDroid 2020" dataset for their work. They performed data preprocessing, including missing value handling, categorical encoding, Boolean mapping, variance threshold feature reduction, feature concatenation, and L2 normalization. They then optimized it with three techniques: original, principal component analysis (PCA) and linear discriminant analysis (LDA). The 'original' approach used "raw vectors", while PCA "retained components" explaining 95% of variance and LDA "projected onto class separating axes" [28]. The results from the synthetic data augmentation technique, presented herein, are compared to Abedin and Mehrub's results in Table 10.

**Table 10.** Comparison to the approaches used by Abedin and Mehrub [28].

| | | **Accuracy** | **Precision** | **Recall** | **F1** |
|---|---|---|---|---|---|
| Original [28] | Average | 93.97 | 93.16 | 93.05 | 93.05 |
| | Best (XGBoost) | 97.47 | 97.03 | 97.31 | 97.16 |
| PCA [28] | Average | 89.48 | 87.68 | 87.37 | 87.46 |
| | Best (ExtraTrees) | 92.16 | 90.33 | 90.53 | 90.37 |
| LDA [28] | Average | 92.00 | 90.95 | 90.56 | 90.73 |
| | Best (ExtraTrees) | 93.25 | 92.19 | 92.01 | 92.09 |
| Sequential model (current) | | 88.58 | 88.66 | 88.58 | 88.54 |
| Random forest (current) | | 95.3 | 95.34 | 95.3 | 95.29 |
| XGBoost (current) | | 95.6 | 95.61 | 95.6 | 95.6 |

The random forest and XGBoost approaches, using the synthetic data augmentation approach presented herein, outperform the average of the techniques used by [28]; however, the best performing technique, XGBoost, using the 'original' pre-processing approach, outperforms all three classifiers presented herein. Given that both [28] and the synthetic data augmentation approach presented herein enhance the performance of the XGBoost technique, it may be possible to combine the two. This is a topic for potential future work.

## 6. Comparison to Other Synthetic Data Approaches

This section compares the approach presented herein, using the three previously discussed classifiers, to the performance of other approaches that used synthetic data. Comparative results from a paper that used the "CICMalDroid 2020" dataset are discussed first. Then results from another paper, using a related dataset, "CCCS-CIC-AndMal2020", are discussed.

Bortoli, Duarte and Guerreiro [14] used SMOTE oversampling to address class imbalance. SMOTE generated synthetic minority class samples by interpolating existing samples. They compared each machine learning model with and without SMOTE, within the training pipeline. The results showed that SMOTE often reduced performance or provided only minor improvements. Table 11 presents the comparative results for the approaches presented in [14].

**Table 11.** Comparison to the Synthetic Data Approaches in [14].

| Approach | Weighted recall |
|---|---|
| XGBoost (F-test selection), non-aggressive, SMOTE (from [14]) | 95.40 |
| XGBoost (F-test selection), non-aggressive, no SMOTE (from [14]) | 95.32 |
| Sequential model | 88.58 |
| Random forest | 95.30 |
| XGBoost | 95.60 |

Bortoli, Duarte and Guerreiro [14] presented the results from 48 approaches to solving the malware classification problem, using machine learning-based classifiers. The performance of these ranged from 34.88% weighted recall – for the BernoulliNB (Chi-square selection), using SMOTE, technique – to 95.40% weighted recall – for the XGBoost (F-test selection), non-aggressive, SMOTE technique. The results presented in Table 12 are for the two best performing techniques from [14]. The XGBoost approach, presented herein, outperforms the two listed best performing techniques from that paper.

While not directly comparable, due to using an alternate dataset, Othman, et al. [15] used CGAN-based synthetic data generation. A conditional GAN was trained on Android malware dynamic features and class labels were used as conditions. The study evaluated two augmentation strategies. The first was multi class generation, where all 12 malware classes were generated at once. The second was binary (or pairwise) generation, where two classes of similar size were generated at one time and later combined. The synthetic data was then mixed with the original data and evaluated using the random forest and LightGBM classifiers.

All three approaches presented herein outperform the best performing approaches from [15]; however, because of a larger number of classification groups in [15] and the use of a different dataset, the two cannot be directly compared. This comparison is a topic for potential future work. Table 12 presents the results for the approaches presented in [15].

**Table 12.** Comparison to the Synthetic Data Approaches in [15].

| Approach | Accuracy | Precision | Recall | F1-score |
|---|---|---|---|---|
| Random Forest on original dataset (from [15]) | 86.00 | 86.45 | 85.75 | 85.75 |
| LGBM after binary-class CGAN generation (from [15]) | 85.28 | 85.07 | 85.28 | 84.81 |

## 7. Conclusion

This study used a VAE-based generative model to generate synthetic malware samples to enhance the effectiveness of malware detection systems. The work presented herein has demonstrated the benefits of combining synthetic data, generated by VAEs, and real-world datasets. The proposed approach produced an improvement in the performance of several machine learning models, as compared to the results from several prior papers.

This improvement was quantified through the accuracy precision, recall, and F1-score metrics. The proposed technique was used for malware classification into the classes of adware, banking malware, SMS malware, riskware, and benign software. It was evaluated using the random forest, XGboost, and sequential model classifiers. The results from this study demonstrate the potential role of generative AI in addressing the challenge of the scarcity of labeled samples for malware variants.

These results also suggest that the implementation of a hybrid method – that leverages the combined strengths of multiple machine learning models, including a generative model – may provide a more comprehensive and accurate system for malware detection. Notably, the approach proposed herein provided the greatest benefit in some of the areas that [7] performed the worst in, demonstrating the prospective benefit of combining multiple classification approaches.

The results presented in Section 6 show that the combination of synthetic data augmentation and the classifier technique that is used have a notable impact on performance. The approaches presented herein compare favorably to those presented in [14] and XGBoost outperformed the approaches presented in [15]. This shows the importance of technique implementation, beyond the basic synthetic data augmentation concept.

Adopting SHAP for interpreting machine learning models was shown, in [7], to provide insight into model behavior. SHAP is based on Shapley values and estimates the contribution of each feature by evaluating its impact across different combinations of input features [7]. This interpretability mechanism helps explain model predictions and identify the features that most greatly influence malware classification. The results produced using the approach presented herein can be analyzed using SHAP, as was demonstrated in [7], to provide these explainability benefits. The proposed approach, of synthetic data generation and use, also provides enhanced classifier performance – in terms of the accuracy, precision, recall and f1-score metrics – as compared to [7]. Analysis of the comparative efficacy of explainability techniques, such as SHAP, with the classifier techniques presented herein (and in [7]), in [8] and in [9] remains a key topic for future work.

The application of generative AI for generating synthetic malware samples is a response to several key challenges in malware detection and analysis. These challenges include the limited availability of high-quality and publicly accessible malware datasets, class imbalance between benign and malicious samples, and the rapid evolution of malware – through techniques such as polymorphism and obfuscation – which make traditional detection methods less effective. Generative models can help address these issues by producing diverse and representative synthetic samples, improving model training and generalization. Future research, focusing on hybrid methodologies and enhanced interpretability, will be crucial to fully harnessing the capabilities of these technologies.

**Acknowledgements**

M. Alharbi used generative AI to assist in paraphrasing in limited instances. All other aspects of the paper preparation, research and analysis were conducted by the authors.

**References**

[1] S. Patil et al., "Improving the robustness of AI-based malware detection using adversarial machine learning," Algorithms, vol. 14, no. 10, p. 297, Oct. 2021, doi: 10.3390/a14100297.
[2] IBM, "Cost of a Data Breach Report," IBM, 2025. [Online]. Available: https://www.ibm.com/reports/data-breach. Accessed: May 6, 2026.
[3] M. Gupta, C. Akiri, K. Aryal, E. Parker, and L. Praharaj, "From ChatGPT to ThreatGPT: Impact of generative AI in cybersecurity and privacy," IEEE Access, vol. 11, pp. 80218–80245, 2023, doi: 10.1109/ACCESS.2023.3300381.
[4] S. Alemohammad, A. I. Humayun, S. Agarwal, J. Collomosse, and R. Baraniuk, "Self-improving diffusion models with synthetic data," arXiv preprint arXiv:2408.16333, Aug. 2024. [Online]. Available: https://arxiv.org/abs/2408.16333. Accessed: May 7, 2026.

[5] Z. Wan, Y. Zhang, and H. He, "Variational autoencoder based synthetic data generation for imbalanced learning," in Proc. 2017 IEEE Symposium Series on Computational Intelligence (SSCI), Honolulu, HI, USA, 2017, pp. 1–7, doi: 10.1109/SSCI.2017.8285168.
[6] T. Liu, J. Fan, G. Li, N. Tang, and X. Du, "Tabular data synthesis with generative adversarial networks: Design space and optimizations," The VLDB Journal, 2023, doi: 10.1007/s00778-023-00807-y.
[7] R. Alenezi and S. A. Ludwig, "Explainability of cybersecurity threats data using SHAP," in Proc. 2021 IEEE Symposium Series on Computational Intelligence (SSCI), Orlando, FL, USA, 2021, pp. 1–10, doi: 10.1109/SSCI50451.2021.9659888.
[8] S. Mahdavifar, A. F. A. Kadir, R. Fatemi, D. Alhadidi, and A. A. Ghorbani, "Dynamic Android malware category classification using semi-supervised deep learning," in Proc. 2020 IEEE Intl. Conf. on Dependable, Autonomic and Secure Computing, Intl. Conf. on Pervasive Intelligence and Computing, Intl. Conf. on Cloud and Big Data Computing, and Intl. Conf. on Cyber Science and Technology Congress (DASC/PiCom/CBDCom/CyberSciTech), 2020, pp. 515–522. [Online]. Available: https://ieeexplore.ieee.org/abstract/document/9251198. Accessed: May 6, 2026.
[9] S. Mahdavifar, D. Alhadidi, and A. A. Ghorbani, "Effective and efficient hybrid Android malware classification using pseudo-label stacked auto-encoder," Journal of Network and Systems Management, vol. 30, no. 1, Art. no. 22, 2022, doi: 10.1007/s10922-021-09634-4.
[10] S. Saad, W. Briguglio, and H. Elmiligi, "The curious case of machine learning in malware detection," arXiv preprint arXiv:1905.07573, 2019. [Online]. Available: https://arxiv.org/abs/1905.07573. Accessed: May 6, 2026.
[11] J. Singh and J. Singh, "A survey on machine learning-based malware detection in executable files," Journal of Systems Architecture, vol. 112, p. 101861, 2021, doi: 10.1016/j.sysarc.2020.101861.
[12] H. Faisal, H. Hindy, S. Gaber, and A. Salem, "A survey on artificial intelligence techniques for malware detection," in Artificial Intelligence, Soft Computing and Applications, 2022.
[13] V. A. Fajardo et al., "On oversampling imbalanced data with deep conditional generative models," Expert Systems with Applications, vol. 169, p. 114463, May 2021, doi: 10.1016/j.eswa.2020.114463.
[14] A. A. Bortoli, D. F. Duarte, and M. R. M. Guerreiro, "Empirical evaluation of SMOTE in Android malware detection with machine learning: Challenges and performance in CICMalDroid 2020," arXiv preprint arXiv:2602.08744, Feb. 2026, doi: 10.48550/arXiv.2602.08744.
[15] F. M. H. Othman et al., "Data augmentation using conditional generative adversarial network (CGAN) for Android malware binary and multi-class classification," in Proc. 2025 IEEE 15th Annual Computing and Communication Workshop and Conference (CCWC), Jan. 2025, pp. 584–592, doi: 10.1109/CCWC62904.2025.10903749.
[16] C. Doersch, "Tutorial on variational autoencoders," arXiv preprint arXiv:1606.05908, 2016. [Online]. Available: https://arxiv.org/abs/1606.05908. Accessed: May 6, 2026.
[17] R. Burks, K. A. Islam, Y. Lu, and J. Li, "Data augmentation with generative models for improved malware detection: A comparative study," in Proc. 2019 IEEE 10th Annual Ubiquitous Computing, Electronics & Mobile Communication Conference (UEMCON), New York, NY, USA, 2019, pp. 660–665, doi: 10.1109/UEMCON47517.2019.8993085.
[18] S. K. Lakshmanaprabu, K. Shankar, M. Ilayaraja, A. W. Nasir, V. Vijayakumar, and N. Chilamkurti, "Random forest for big data classification in the Internet of Things using optimal features," International Journal of Machine Learning and Cybernetics, vol. 10, no. 10, pp. 2609–2618, 2019, doi: 10.1007/s13042-018-00916-z.
[19] S. M. Lundberg, G. G. Erion, and S.-I. Lee, "Consistent individualized feature attribution for tree ensembles," arXiv preprint arXiv:1802.03888, 2018. [Online]. Available: https://arxiv.org/abs/1802.03888. Accessed: May 6, 2026.
[20] L. Khaidem, S. Saha, and S. Roy Dey, "Predicting the direction of stock market prices using random forest," arXiv preprint arXiv:1605.00003, Apr. 2016. [Online]. Available: https://arxiv.org/abs/1605.00003. Accessed: May 6, 2026.

[21] R. Kumar and G. S., "Malware classification using XGBoost-gradient boosted decision tree," Advances in Science, Technology and Engineering Systems Journal, vol. 5, no. 5, pp. 536–549, 2020, doi: 10.25046/aj050566.
[22] T. Chen and C. Guestrin, "XGBoost: A scalable tree boosting system," in Proc. 22nd ACM SIGKDD International Conference on Knowledge Discovery and Data Mining, San Francisco, CA, USA, 2016, pp. 785–794, doi: 10.1145/2939672.2939785.
[23] A. Gulli, A. Kapoor, and S. Pal, Deep Learning with TensorFlow 2 and Keras. Birmingham, U.K.: Packt Publishing, 2019.
[24] Canadian Institute for Cybersecurity, "Datasets," University of New Brunswick. [Online]. Available: https://www.unb.ca/cic/. Accessed: May 6, 2026.
[25] XGBoost Developers, "XGBoost parameters," XGBoost Documentation. [Online]. Available: https://xgboost.readthedocs.io/en/stable/parameter.html. Accessed: May 6, 2026.
[26] XGBoost Developers, "Tree methods," XGBoost Documentation. [Online]. Available: https://xgboost.readthedocs.io/en/stable/treemethod.html. Accessed: May 8, 2026.
[27] XGBoost Developers, "Python API reference," XGBoost Documentation. [Online]. Available: https://xgboost.readthedocs.io/en/latest/python/python_api.html. Accessed: May 6, 2026.
[28] M. M.-H.-Z. Abedin and T. Mehrub, "Comparison of multiple classifiers for Android malware detection with emphasis on feature insights using CICMalDroid 2020 dataset," in *Proc. 2025 IEEE 7th International Conference on Sustainable Technologies for Industry 5.0 (STI)*, Dhaka, Bangladesh, Dec. 2025, pp. 1–6, doi: 10.1109/STI69347.2025.11367549.